\pdfoutput=1
\documentclass[11pt,twoside,a4paper,pdftex,cmspaper,final,collab]{cms-tdr}

\begin{document}\cmsNoteHeader{CFT-09-025}
%%%%%%%%%%%%%%%%%%%%%%%%%%%%%%%%%%%%%%%%%%%%%%%%%%%%%%%%%%%%%%%%%%%%
%
%  Common definitions
%
%  N.B. use of \providecommand rather than \newcommand means
%       that a definition is ignored if already specified
%
%                                              L. Taylor 18 Feb 2005
%%%%%%%%%%%%%%%%%%%%%%%%%%%%%%%%%%%%%%%%%%%%%%%%%%%%%%%%%%%%%%%%%%%%

%%%%%%%%%%%%%%%%%%%%%%%%%%%%%%%%%%%%%%%%%%%%%%%%%%%%%%%%%%%%%%%%%%%%
%
% Hyphenations (only need to add here if you get a nasty word break)
%
\hyphenation{env-iron-men-tal}%    just an example
\hyphenation{had-ron-i-za-tion}
\hyphenation{cal-or-i-me-ter}
\hyphenation{de-vices}
%
% Hyphenations-end
%
% CVS info. These are modified by cvs at checkout time.
% The last version of these macros found before the maketitle will be the one on the front page,
% so only the main file is tracked.
% Do not edit by hand!
\RCS$Revision: 1.27 $
% \RCS$Date: 2010/01/11 22:51:12 $
\RCS$Date: 2010/01/11 22:51:12 $
\RCS$Name:  $
%%%%%%%%%%%%% ptdr definitions %%%%%%%%%%%%%%%%%%%%%
%%%%%%%%%%%%%%%%%%%%%%%%%%%%%%%%%%%%%%%%%%%%%%%%%%%%%%%%%%%%%%%%%%%%
%
%  Common definitions
%
%  N.B. use of \providecommand rather than \newcommand means
%       that a definition is ignored if already specified
%
%                                              L. Taylor 18 Feb 2005
%%%%%%%%%%%%%%%%%%%%%%%%%%%%%%%%%%%%%%%%%%%%%%%%%%%%%%%%%%%%%%%%%%%%

% Some shorthand
% turn off italics
\newcommand {\etal}{\mbox{et al.}\xspace} %et al. - no preceding comma
\newcommand {\ie}{\mbox{i.e.}\xspace}     %i.e.
\newcommand {\eg}{\mbox{e.g.}\xspace}     %e.g.
\newcommand {\etc}{\mbox{etc.}\xspace}     %etc.
\newcommand {\vs}{\mbox{\sl vs.}\xspace}      %vs.
\newcommand {\mdash}{\ensuremath{\mathrm{-}}} % for use within formulas

% some terms whose definition we may change
\newcommand {\Lone}{Level-1\xspace} % Level-1 or L1 ?
\newcommand {\Ltwo}{Level-2\xspace}
\newcommand {\Lthree}{Level-3\xspace}

% Some software programs (alphabetized)
\providecommand{\ACERMC} {\textsc{AcerMC}\xspace}
\providecommand{\ALPGEN} {{\textsc{alpgen}}\xspace}
\providecommand{\CHARYBDIS} {{\textsc{charybdis}}\xspace}
\providecommand{\CMKIN} {\textsc{cmkin}\xspace}
\providecommand{\CMSIM} {{\textsc{cmsim}}\xspace}
\providecommand{\CMSSW} {{\textsc{cmssw}}\xspace}
\providecommand{\COBRA} {{\textsc{cobra}}\xspace}
\providecommand{\COCOA} {{\textsc{cocoa}}\xspace}
\providecommand{\COMPHEP} {\textsc{CompHEP}\xspace}
\providecommand{\EVTGEN} {{\textsc{evtgen}}\xspace}
\providecommand{\FAMOS} {{\textsc{famos}}\xspace}
\providecommand{\GARCON} {\textsc{garcon}\xspace}
\providecommand{\GARFIELD} {{\textsc{garfield}}\xspace}
\providecommand{\GEANE} {{\textsc{geane}}\xspace}
\providecommand{\GEANTfour} {{\textsc{geant4}}\xspace}
\providecommand{\GEANTthree} {{\textsc{geant3}}\xspace}
\providecommand{\GEANT} {{\textsc{geant}}\xspace}
\providecommand{\HDECAY} {\textsc{hdecay}\xspace}
\providecommand{\HERWIG} {{\textsc{herwig}}\xspace}
\providecommand{\HIGLU} {{\textsc{higlu}}\xspace}
\providecommand{\HIJING} {{\textsc{hijing}}\xspace}
\providecommand{\IGUANA} {\textsc{iguana}\xspace}
\providecommand{\ISAJET} {{\textsc{isajet}}\xspace}
\providecommand{\ISAPYTHIA} {{\textsc{isapythia}}\xspace}
\providecommand{\ISASUGRA} {{\textsc{isasugra}}\xspace}
\providecommand{\ISASUSY} {{\textsc{isasusy}}\xspace}
\providecommand{\ISAWIG} {{\textsc{isawig}}\xspace}
\providecommand{\MADGRAPH} {\textsc{MadGraph}\xspace}
\providecommand{\MCATNLO} {\textsc{mc@nlo}\xspace}
\providecommand{\MCFM} {\textsc{mcfm}\xspace}
\providecommand{\MILLEPEDE} {{\textsc{millepede}}\xspace}
\providecommand{\ORCA} {{\textsc{orca}}\xspace}
\providecommand{\OSCAR} {{\textsc{oscar}}\xspace}
\providecommand{\PHOTOS} {\textsc{photos}\xspace}
\providecommand{\PROSPINO} {\textsc{prospino}\xspace}
\providecommand{\PYTHIA} {{\textsc{pythia}}\xspace}
\providecommand{\SHERPA} {{\textsc{sherpa}}\xspace}
\providecommand{\TAUOLA} {\textsc{tauola}\xspace}
\providecommand{\TOPREX} {\textsc{TopReX}\xspace}
\providecommand{\XDAQ} {{\textsc{xdaq}}\xspace}

%  Experiments
\newcommand {\DZERO}{D\O\xspace}     %etc.

% Measurements and units...

\newcommand{\de}{\ensuremath{^\circ}}
\newcommand{\ten}[1]{\ensuremath{\times \text{10}^\text{#1}}}
\newcommand{\unit}[1]{\ensuremath{\text{\,#1}}\xspace}
\newcommand{\mum}{\ensuremath{\,\mu\text{m}}\xspace}
\newcommand{\micron}{\ensuremath{\,\mu\text{m}}\xspace}
\newcommand{\cm}{\ensuremath{\,\text{cm}}\xspace}
\newcommand{\mm}{\ensuremath{\,\text{mm}}\xspace}
\newcommand{\mus}{\ensuremath{\,\mu\text{s}}\xspace}
\newcommand{\keV}{\ensuremath{\,\text{ke\hspace{-.08em}V}}\xspace}
\newcommand{\MeV}{\ensuremath{\,\text{Me\hspace{-.08em}V}}\xspace}
\newcommand{\GeV}{\ensuremath{\,\text{Ge\hspace{-.08em}V}}\xspace}
\newcommand{\TeV}{\ensuremath{\,\text{Te\hspace{-.08em}V}}\xspace}
\newcommand{\PeV}{\ensuremath{\,\text{Pe\hspace{-.08em}V}}\xspace}
\newcommand{\keVc}{\ensuremath{{\,\text{ke\hspace{-.08em}V\hspace{-0.16em}/\hspace{-0.08em}c}}}\xspace}
\newcommand{\MeVc}{\ensuremath{{\,\text{Me\hspace{-.08em}V\hspace{-0.16em}/\hspace{-0.08em}c}}}\xspace}
\newcommand{\GeVc}{\ensuremath{{\,\text{Ge\hspace{-.08em}V\hspace{-0.16em}/\hspace{-0.08em}c}}}\xspace}
\newcommand{\TeVc}{\ensuremath{{\,\text{Te\hspace{-.08em}V\hspace{-0.16em}/\hspace{-0.08em}c}}}\xspace}
\newcommand{\keVcc}{\ensuremath{{\,\text{ke\hspace{-.08em}V\hspace{-0.16em}/\hspace{-0.08em}c}^\text{2}}}\xspace}
\newcommand{\MeVcc}{\ensuremath{{\,\text{Me\hspace{-.08em}V\hspace{-0.16em}/\hspace{-0.08em}c}^\text{2}}}\xspace}
\newcommand{\GeVcc}{\ensuremath{{\,\text{Ge\hspace{-.08em}V\hspace{-0.16em}/\hspace{-0.08em}c}^\text{2}}}\xspace}
\newcommand{\TeVcc}{\ensuremath{{\,\text{Te\hspace{-.08em}V\hspace{-0.16em}/\hspace{-0.08em}c}^\text{2}}}\xspace}

\newcommand{\pbinv} {\mbox{\ensuremath{\,\text{pb}^\text{$-$1}}}\xspace}
\newcommand{\fbinv} {\mbox{\ensuremath{\,\text{fb}^\text{$-$1}}}\xspace}
\newcommand{\nbinv} {\mbox{\ensuremath{\,\text{nb}^\text{$-$1}}}\xspace}
\newcommand{\percms}{\ensuremath{\,\text{cm}^\text{$-$2}\,\text{s}^\text{$-$1}}\xspace}
\newcommand{\lumi}{\ensuremath{\mathcal{L}}\xspace}
\newcommand{\Lumi}{\ensuremath{\mathcal{L}}\xspace}%both upper and lower
%
% Need a convention here:
\newcommand{\LvLow}  {\ensuremath{\mathcal{L}=\text{10}^\text{32}\,\text{cm}^\text{$-$2}\,\text{s}^\text{$-$1}}\xspace}
\newcommand{\LLow}   {\ensuremath{\mathcal{L}=\text{10}^\text{33}\,\text{cm}^\text{$-$2}\,\text{s}^\text{$-$1}}\xspace}
\newcommand{\lowlumi}{\ensuremath{\mathcal{L}=\text{2}\times \text{10}^\text{33}\,\text{cm}^\text{$-$2}\,\text{s}^\text{$-$1}}\xspace}
\newcommand{\LMed}   {\ensuremath{\mathcal{L}=\text{2}\times \text{10}^\text{33}\,\text{cm}^\text{$-$2}\,\text{s}^\text{$-$1}}\xspace}
\newcommand{\LHigh}  {\ensuremath{\mathcal{L}=\text{10}^\text{34}\,\text{cm}^\text{$-$2}\,\text{s}^\text{$-$1}}\xspace}
\newcommand{\hilumi} {\ensuremath{\mathcal{L}=\text{10}^\text{34}\,\text{cm}^\text{$-$2}\,\text{s}^\text{$-$1}}\xspace}

% Some usual physics terms

\newcommand{\zp}{\ensuremath{\mathrm{Z}^\prime}\xspace}

% SM (still to be classified)

\newcommand{\kt}{\ensuremath{k_{\mathrm{T}}}\xspace}
\newcommand{\BC}{\ensuremath{{B_{\mathrm{c}}}}\xspace}
\newcommand{\bbarc}{\ensuremath{{\overline{b}c}}\xspace}
\newcommand{\bbbar}{\ensuremath{{b\overline{b}}}\xspace}
\newcommand{\ccbar}{\ensuremath{{c\overline{c}}}\xspace}
\newcommand{\JPsi}{\ensuremath{{J}/\psi}\xspace}
\newcommand{\bspsiphi}{\ensuremath{B_s \to \JPsi\, \phi}\xspace}
\newcommand{\AFB}{\ensuremath{A_\mathrm{FB}}\xspace}
\newcommand{\EE}{\ensuremath{e^+e^-}\xspace}
\newcommand{\MM}{\ensuremath{\mu^+\mu^-}\xspace}
\newcommand{\TT}{\ensuremath{\tau^+\tau^-}\xspace}
\newcommand{\wangle}{\ensuremath{\sin^{2}\theta_{\mathrm{eff}}^\mathrm{lept}(M^2_\mathrm{Z})}\xspace}
\newcommand{\ttbar}{\ensuremath{{t\overline{t}}}\xspace}
\newcommand{\stat}{\ensuremath{\,\text{(stat.)}}\xspace}
\newcommand{\syst}{\ensuremath{\,\text{(syst.)}}\xspace}
% these moved to similar defs
%\newcommand{\Etmiss}{\ensuremath{E_{\mathrm{T}\!{\rm miss}}}}
%\newcommand{\VEtmiss}{\ensuremath{{\vec E}_{\mathrm{T}\!{\rm miss}}}}

%%%  E-gamma definitions
\newcommand{\HGG}{\ensuremath{\mathrm{H}\to\gamma\gamma}}
\newcommand{\gev}{\GeV}
\newcommand{\GAMJET}{\ensuremath{\gamma + \mathrm{jet}}}
\newcommand{\PPTOJETS}{\ensuremath{\mathrm{pp}\to\mathrm{jets}}}
\newcommand{\PPTOGG}{\ensuremath{\mathrm{pp}\to\gamma\gamma}}
\newcommand{\PPTOGAMJET}{\ensuremath{\mathrm{pp}\to\gamma +
\mathrm{jet}
}}
\newcommand{\MH}{\ensuremath{\mathrm{M_{\mathrm{H}}}}}
\newcommand{\RNINE}{\ensuremath{\mathrm{R}_\mathrm{9}}}
\newcommand{\DR}{\ensuremath{\Delta\mathrm{R}}}

% Physics symbols ...

\newcommand{\PT}{\ensuremath{p_{\mathrm{T}}}\xspace}
\newcommand{\pt}{\ensuremath{p_{\mathrm{T}}}\xspace}
\newcommand{\ET}{\ensuremath{E_{\mathrm{T}}}\xspace}
\newcommand{\HT}{\ensuremath{H_{\mathrm{T}}}\xspace}
\newcommand{\et}{\ensuremath{E_{\mathrm{T}}}\xspace}
\newcommand{\Em}{\ensuremath{E\!\!\!/}\xspace}
\newcommand{\Pm}{\ensuremath{p\!\!\!/}\xspace}
\newcommand{\PTm}{\ensuremath{{p\!\!\!/}_{\mathrm{T}}}\xspace}
\newcommand{\ETm}{\ensuremath{E_{\mathrm{T}}^{\mathrm{miss}}}\xspace}
\newcommand{\MET}{\ensuremath{E_{\mathrm{T}}^{\mathrm{miss}}}\xspace}
\newcommand{\ETmiss}{\ensuremath{E_{\mathrm{T}}^{\mathrm{miss}}}\xspace}
\newcommand{\VEtmiss}{\ensuremath{{\vec E}_{\mathrm{T}}^{\mathrm{miss}}}\xspace}

%%%%%%
% From Albert
%

\newcommand{\ga}{\ensuremath{\gtrsim}}
\newcommand{\la}{\ensuremath{\lesssim}}
\newcommand{\swsq}{\ensuremath{\sin^2\theta_W}\xspace}
\newcommand{\cwsq}{\ensuremath{\cos^2\theta_W}\xspace}
\newcommand{\tanb}{\ensuremath{\tan\beta}\xspace}
\newcommand{\tanbsq}{\ensuremath{\tan^{2}\beta}\xspace}
\newcommand{\sidb}{\ensuremath{\sin 2\beta}\xspace}
\newcommand{\alpS}{\ensuremath{\alpha_S}\xspace}
\newcommand{\alpt}{\ensuremath{\tilde{\alpha}}\xspace}

\newcommand{\QL}{\ensuremath{Q_L}\xspace}
\newcommand{\sQ}{\ensuremath{\tilde{Q}}\xspace}
\newcommand{\sQL}{\ensuremath{\tilde{Q}_L}\xspace}
\newcommand{\ULC}{\ensuremath{U_L^C}\xspace}
\newcommand{\sUC}{\ensuremath{\tilde{U}^C}\xspace}
\newcommand{\sULC}{\ensuremath{\tilde{U}_L^C}\xspace}
\newcommand{\DLC}{\ensuremath{D_L^C}\xspace}
\newcommand{\sDC}{\ensuremath{\tilde{D}^C}\xspace}
\newcommand{\sDLC}{\ensuremath{\tilde{D}_L^C}\xspace}
\newcommand{\LL}{\ensuremath{L_L}\xspace}
\newcommand{\sL}{\ensuremath{\tilde{L}}\xspace}
\newcommand{\sLL}{\ensuremath{\tilde{L}_L}\xspace}
\newcommand{\ELC}{\ensuremath{E_L^C}\xspace}
\newcommand{\sEC}{\ensuremath{\tilde{E}^C}\xspace}
\newcommand{\sELC}{\ensuremath{\tilde{E}_L^C}\xspace}
\newcommand{\sEL}{\ensuremath{\tilde{E}_L}\xspace}
\newcommand{\sER}{\ensuremath{\tilde{E}_R}\xspace}
\newcommand{\sFer}{\ensuremath{\tilde{f}}\xspace}
\newcommand{\sQua}{\ensuremath{\tilde{q}}\xspace}
\newcommand{\sUp}{\ensuremath{\tilde{u}}\xspace}
\newcommand{\suL}{\ensuremath{\tilde{u}_L}\xspace}
\newcommand{\suR}{\ensuremath{\tilde{u}_R}\xspace}
\newcommand{\sDw}{\ensuremath{\tilde{d}}\xspace}
\newcommand{\sdL}{\ensuremath{\tilde{d}_L}\xspace}
\newcommand{\sdR}{\ensuremath{\tilde{d}_R}\xspace}
\newcommand{\sTop}{\ensuremath{\tilde{t}}\xspace}
\newcommand{\stL}{\ensuremath{\tilde{t}_L}\xspace}
\newcommand{\stR}{\ensuremath{\tilde{t}_R}\xspace}
\newcommand{\stone}{\ensuremath{\tilde{t}_1}\xspace}
\newcommand{\sttwo}{\ensuremath{\tilde{t}_2}\xspace}
\newcommand{\sBot}{\ensuremath{\tilde{b}}\xspace}
\newcommand{\sbL}{\ensuremath{\tilde{b}_L}\xspace}
\newcommand{\sbR}{\ensuremath{\tilde{b}_R}\xspace}
\newcommand{\sbone}{\ensuremath{\tilde{b}_1}\xspace}
\newcommand{\sbtwo}{\ensuremath{\tilde{b}_2}\xspace}
\newcommand{\sLep}{\ensuremath{\tilde{l}}\xspace}
\newcommand{\sLepC}{\ensuremath{\tilde{l}^C}\xspace}
\newcommand{\sEl}{\ensuremath{\tilde{e}}\xspace}
\newcommand{\sElC}{\ensuremath{\tilde{e}^C}\xspace}
\newcommand{\seL}{\ensuremath{\tilde{e}_L}\xspace}
\newcommand{\seR}{\ensuremath{\tilde{e}_R}\xspace}
\newcommand{\snL}{\ensuremath{\tilde{\nu}_L}\xspace}
\newcommand{\sMu}{\ensuremath{\tilde{\mu}}\xspace}
\newcommand{\sNu}{\ensuremath{\tilde{\nu}}\xspace}
\newcommand{\sTau}{\ensuremath{\tilde{\tau}}\xspace}
\newcommand{\Glu}{\ensuremath{g}\xspace}
\newcommand{\sGlu}{\ensuremath{\tilde{g}}\xspace}
\newcommand{\Wpm}{\ensuremath{W^{\pm}}\xspace}
\newcommand{\sWpm}{\ensuremath{\tilde{W}^{\pm}}\xspace}
\newcommand{\Wz}{\ensuremath{W^{0}}\xspace}
\newcommand{\sWz}{\ensuremath{\tilde{W}^{0}}\xspace}
\newcommand{\sWino}{\ensuremath{\tilde{W}}\xspace}
\newcommand{\Bz}{\ensuremath{B^{0}}\xspace}
\newcommand{\sBz}{\ensuremath{\tilde{B}^{0}}\xspace}
\newcommand{\sBino}{\ensuremath{\tilde{B}}\xspace}
\newcommand{\Zz}{\ensuremath{Z^{0}}\xspace}
\newcommand{\sZino}{\ensuremath{\tilde{Z}^{0}}\xspace}
\newcommand{\sGam}{\ensuremath{\tilde{\gamma}}\xspace}
\newcommand{\chiz}{\ensuremath{\tilde{\chi}^{0}}\xspace}
\newcommand{\chip}{\ensuremath{\tilde{\chi}^{+}}\xspace}
\newcommand{\chim}{\ensuremath{\tilde{\chi}^{-}}\xspace}
\newcommand{\chipm}{\ensuremath{\tilde{\chi}^{\pm}}\xspace}
\newcommand{\Hone}{\ensuremath{H_{d}}\xspace}
\newcommand{\sHone}{\ensuremath{\tilde{H}_{d}}\xspace}
\newcommand{\Htwo}{\ensuremath{H_{u}}\xspace}
\newcommand{\sHtwo}{\ensuremath{\tilde{H}_{u}}\xspace}
\newcommand{\sHig}{\ensuremath{\tilde{H}}\xspace}
\newcommand{\sHa}{\ensuremath{\tilde{H}_{a}}\xspace}
\newcommand{\sHb}{\ensuremath{\tilde{H}_{b}}\xspace}
\newcommand{\sHpm}{\ensuremath{\tilde{H}^{\pm}}\xspace}
\newcommand{\hz}{\ensuremath{h^{0}}\xspace}
\newcommand{\Hz}{\ensuremath{H^{0}}\xspace}
\newcommand{\Az}{\ensuremath{A^{0}}\xspace}
\newcommand{\Hpm}{\ensuremath{H^{\pm}}\xspace}
\newcommand{\sGra}{\ensuremath{\tilde{G}}\xspace}
\newcommand{\mtil}{\ensuremath{\tilde{m}}\xspace}
\newcommand{\rpv}{\ensuremath{\rlap{\kern.2em/}R}\xspace}
\newcommand{\LLE}{\ensuremath{LL\bar{E}}\xspace}
\newcommand{\LQD}{\ensuremath{LQ\bar{D}}\xspace}
\newcommand{\UDD}{\ensuremath{\overline{UDD}}\xspace}
\newcommand{\Lam}{\ensuremath{\lambda}\xspace}
\newcommand{\Lamp}{\ensuremath{\lambda'}\xspace}
\newcommand{\Lampp}{\ensuremath{\lambda''}\xspace}
\newcommand{\spinbd}[2]{\ensuremath{\bar{#1}_{\dot{#2}}}\xspace}

\newcommand{\MD}{\ensuremath{{M_\mathrm{D}}}\xspace}% ED mass
\newcommand{\Mpl}{\ensuremath{{M_\mathrm{Pl}}}\xspace}% Planck mass
\newcommand{\Rinv} {\ensuremath{{R}^{-1}}\xspace}

%%%%%%%%%%%%%%%%%%%%%%%%%%%%%%%%%%%%%%%%%%%%%%%%%%%%%%%%%%%%%%%%%%%%
%
% Hyphenations (only need to add here if you get a nasty word break)
%
\hyphenation{en-viron-men-tal}%    just an example

%%%%%%%%%%%%%%%  Title page %%%%%%%%%%%%%%%%%%%%%%%%
\cmsNoteHeader{CFT-09-025}
\title{Fine Synchronization of the CMS Muon Drift-Tube 
  Local Trigger using Cosmic Rays}% Force line breaks with \\

%Author is always "The CMS Collaboration" for PAS, so author, etc will be ignored
\address[cern]{CERN}
\address[neu]{Northeastern University}
\author[cern]{The CMS Collaboration}

% please supply the date in yyyy/mm/dd format. Today has been
% redefined to do so, but it should be fixed as of the final release date.
\date{\today}

% note that you cannot use \verb in the abstract text
\abstract{

 The CMS experiment uses self-triggering arrays of drift tubes in the barrel muon trigger to 
  perform the identification of the correct bunch crossing. The identification is unique only
   if the trigger chain is correctly synchronized. In this 
  paper, the synchronization performed during an extended cosmic ray run 
  is described and the results are reported. The random arrival time of cosmic ray muons 
  allowed several synchronization aspects to be studied and a simple method for the fine
   synchronization of the Drift Tube Local Trigger at LHC to be developed.

}

% Do not comment out the following hypersetup lines (metadata). They will disappear in NODRAFT mode and are needed by CDS.
% Also: make sure that the values of the metadata items are sensible.
\hypersetup{%
pdfauthor={Anna Meneguzzo},%
pdftitle={Fine Synchronization of the CMS Muon Drift-Tube 
  Local Trigger using Cosmic Rays},%
pdfsubject={CMS},%
pdfkeywords={CMS, physics, trigger}}

\maketitle %maketitle comes after all the front information has been supplied
%%%%%%%%%%%%%%%%%%%%%%%%%%%%%%%%  Begin text %%%%%%%%%%%%%%%%%%%%%%%%%%%%%
\section{Introduction}

The primary goal of the Compact Muon Solenoid (CMS) experiment~\cite{CMSall}
is to explore physics at the TeV energy scale, exploiting the proton-proton collisions
delivered by the Large Hadron Collider (LHC)~\cite{LHC}.
In order to achieve very high luminosity, the bunch crossing (BX) frequency of the proton beams is large: $40.08$ MHz.
This large BX frequency requires an accurate synchronization of all the detector components that trigger the data acquisition.

 In the muon barrel detector \cite{Mutdr} 
    the trigger electronics will start the signal processing
     with a variable delay since the time of flight (TOF) of a muon between the interaction point and the detector
      varies from 12 ns for the closest muon
     station to 35 ns for the farthest one.
The Drift Tubes Local Trigger (DTLT) electronics generates trigger  
primitives based on the measurements of the Drift Tubes (DT) detectors  
of the barrel muon system of CMS. In each chamber, the DTLT generates  
track segments from the alignment of hits and assigns the BX. A unique  
identification of the BX, independent of the muon momentum and  
direction, requires the phase of the sampling clock of the DTLT  
electronics to  be adjusted with a precision of 1 ns. This adjustment  
can be achieved using a programmable  delay (0--25 ns) available for
each chamber. This "fine synchronization" is the topic of this paper.

The method initially planned for the LHC operations consists of a scan of the fine delay for each chamber
 to maximize the DTLT efficiency, using muons from LHC p-p collisions. This method, although very
  reliable, has the disadvantage of being time consuming and requiring
   rather stable LHC running conditions.  
In this paper we present an alternative and independent way to synchronize the DT Local Trigger system.
 Cosmic ray muons are asynchronous with respect to the clock and they span the whole 25 ns cycle, replicating the effect of many
    LHC runs with shifted phases of the sampling clock. The DTLT performance is optimal when the muon arrival time in a chamber
has a well defined phase with respect to the local clock of the chamber. Cosmic ray muon data, taken in 2008
 during the Cosmic Run At Four Tesla (CRAFT),
   have been used to measure this optimal phase for
  all chambers, based on the capability of the DT detector
  to measure precisely the track crossing time. These measurements, performed before the LHC 
  start-up, can be used during data taking for a
  fast fine synchronization of the DTLTs. Muons are produced in p-p collisions at a well defined time
  with respect to the LHC clock; the local clock of each chamber will be adjusted such that the mean phase between the 
muon signal from the chamber and the local clock is equal to the optimal phase 
determined with the cosmic ray data analysis.

\phantom{a}

 This paper is organized as follows. Section 2 describes the features of the DTLT, with particular attention to 
 the fine synchronization issues specific to the DTLT electronics. 
  Section 3 describes the method which was used to synchronize the chambers during CRAFT.
   This method relies on the fact that cosmic
  ray  muons arrive at random times with respect to the clock
   and can not be applied to LHC collision data.
  Section 4 describes the measurement of the optimal phase of the DTLT trigger for
  all chambers performed with cosmic ray muon data taken in 2008 during CRAFT.
  Section 5 explains how the results of Section 4 can be used to perform
  the synchronization for normal LHC operation.

\section{The DT local trigger and its synchronization}

A detailed description of the CMS  experiment
can be found elsewhere \cite{CMSall}. The central feature of the CMS apparatus is a 
superconducting solenoid, of 6~m internal diameter, providing a field of 3.8~T.
Within the field volume are the silicon pixel and strip tracker, the
crystal electromagnetic calorimeter and the brass/scintillator
hadron calorimeter.  Muons are measured in gas-ionization detectors
embedded in the steel return yoke. The Trigger and Timing Control system 
(TTC)\cite{TAYLORlhc} distributes the $40.08$~MHz clock and broadcasts the control signals and the general Level 1 (L1) trigger strobe 
to the electronics of  the experiment.

CMS uses a right-handed coordinate system, with the origin at the
nominal collision point, the $x$-axis pointing to the center of the
LHC, the $y$-axis pointing up (perpendicular to the LHC plane), and
the $z$-axis along the anticlockwise-beam direction.
The polar angle, $\theta$, is measured from the positive $z$-axis and
the azimuthal angle, $\phi$, is measured in the $x$-$y$ plane.

 The barrel muon detector \cite{Mutdr} consists of 250 chambers
 arranged in four muon stations named, starting from the interaction point, MB1, MB2, MB3, and MB4, embedded in the 
steel yoke. The steel yoke is divided along the beam direction in five wheels, numbered from -2 to +2, each one with 12 
sectors in the transverse plane, sector 1 being at $\phi$~=~0.
 A chamber is made of layers of drift tubes staggered by half a cell. Layers are arranged in groups
 of four to form SuperLayers (SL). Two SLs measure the track in the
  $r$-$\phi$ plane, while in the three innermost stations a third SL measures
  also the track in the longitudinal plane ($r$-$\theta$ plane). Each DT chamber consists of the three (two in MB4 stations) SLs of drift
   tubes, their Read Out Boards (ROB) electronics, the TRigger Board (TRB) electronics and one  Trigger and Timing Control receiver (TTCrx).
The TTCrx receives the $40.08$ MHz clock from the TTC and broadcasts it both to the ROBs
 and to the TRBs, hence 
    a muon chamber can be considered
    as one intrinsically synchronous block. 
% \begin{figure}[htbp]
\begin{figure}[h]
 \begin{center}
  \resizebox{11cm}{!}{\includegraphics{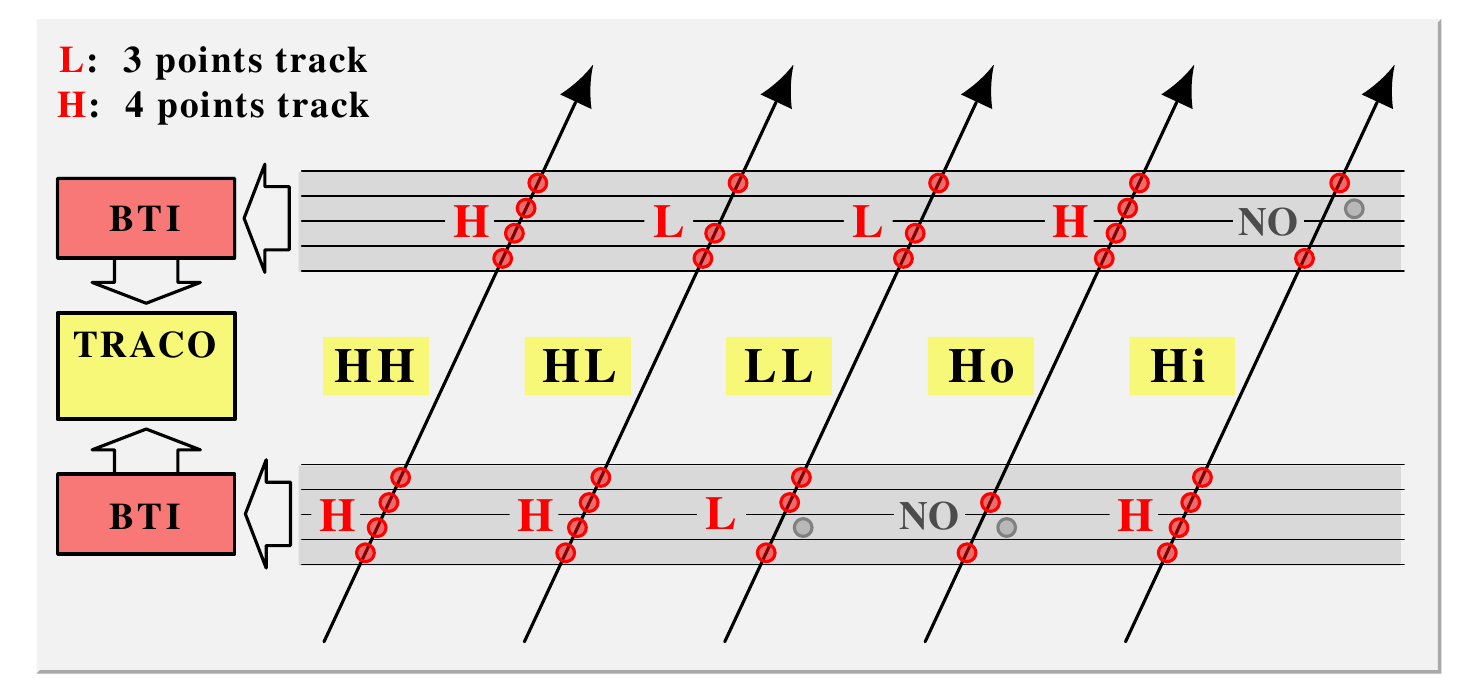}}
  \caption{DT Local Trigger quality classification. Hits along the muon track in the two SLs of the $r$-$\phi$
  view are indicated by small circles; the output of the trigger primitives from a single SL,
 generated by the Bunch and Track Identifiers (BTI),
 and the trigger primitives correlating hits in two SLs, generated by the TRack COrrelator device (TRACO),
 are shown for different cases.
}
  \label{fig1}
 \end{center}
\end{figure}
 %[htbp]
The signals of the DT wires are sampled and processed by the TRB to provide DTLT primitives, i.e.\ 
track segments and their associated BX,
which are sent to the next steps of the muon trigger~\cite{TRIGtdr}. 
 The front-end trigger device of the TRB is the Bunch and Track Identifier (BTI),
  which finds alignments of hits in each SL and assigns them a BX number~\cite{DTtrigAll}.
   The performance depends on two relevant parameters:
      the drift velocity and the time pedestal
      for the drift time computation~\cite{DTbti2000}. 
	    The device samples each wire at twice the LHC frequency.
	      The time of signal sampling de-facto defines the uncertainty
	      in the BTI ``fit'' of the hits and in the accuracy of the track segments parameters.
	      The output of the BTI is provided at 40.08 MHz, selecting the best alignments found in the last two sampling clocks.
  
\begin{figure}[h]
 \begin{center}
  \resizebox{14cm}{!}{\includegraphics{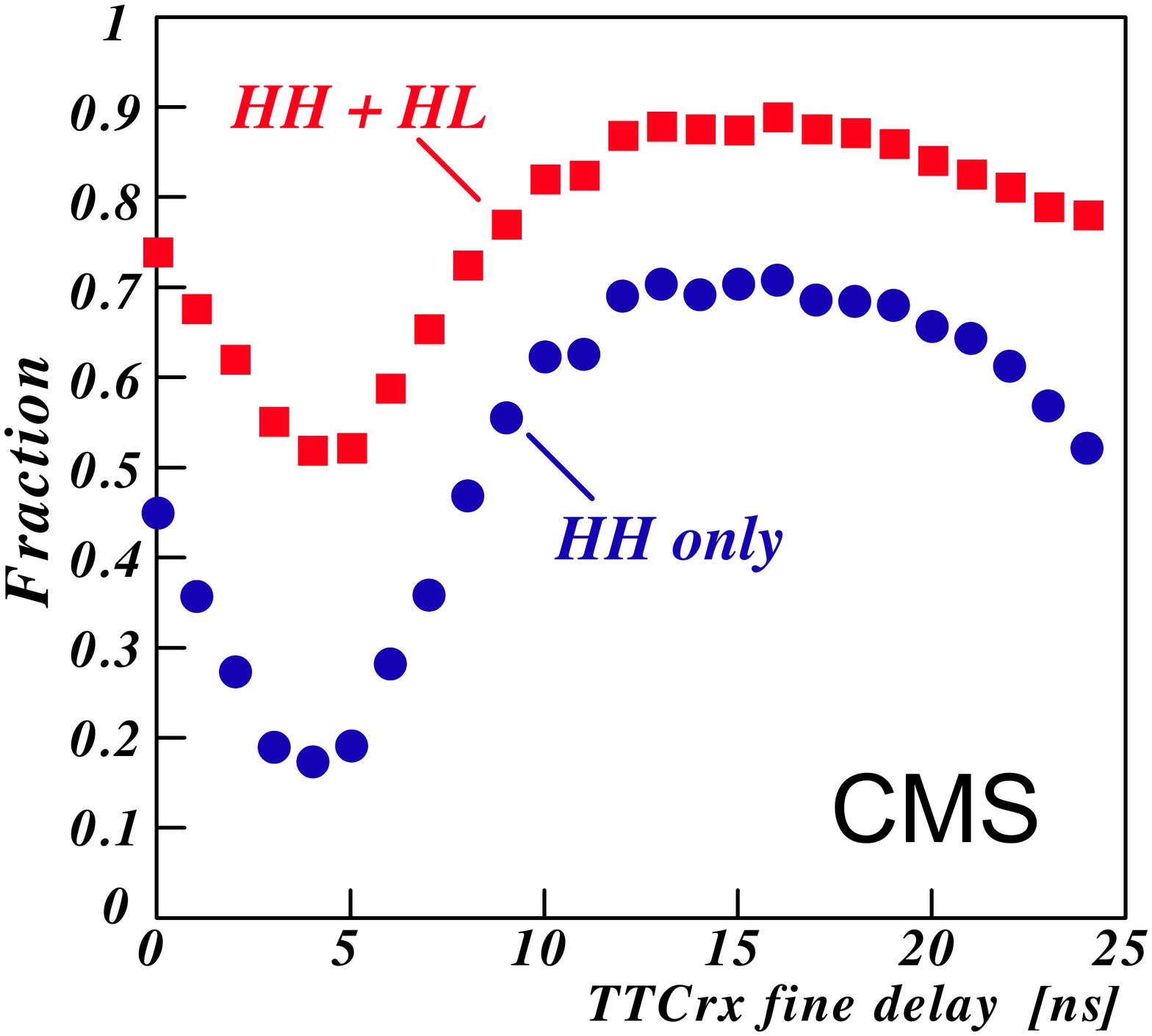}}
  \caption{DTLT fraction for HH and HH+HL trigger primitives with respect to all triggered events as a function of the TTCrx
   delay adjustment. This was measured using a bunched muon test beam and an external scintillator trigger~\cite{DTfineSync}.}
  \label{fig3}
 \end{center}
\end{figure}
All possible trigger primitive types delivered by the DTLT are sketched in Fig.~\ref{fig1}.
 In each SL, the alignments of 3 out of 4, or 4 out of 4 hits are named Low~(L) or High~(H)
 quality, respectively. If such alignments are matched together between the two SLs, the quality of the 
 trigger primitive then becomes HH, HL or LL. The best possible quality (HH) is delivered when hits of
  all the eight layers from the two SLs in the $r$-$\phi$ view are aligned.
Accurate studies were performed with a $\sim$40 MHz bunched muon beam at the CERN SPS in 2003 and 2004 as reported 
in Ref.~\cite{DTtrigg,DTfineSync}.
  The DTLT efficiency was measured as a function of the TTCrx fine delays applied and hence of the time phase of the track
  with respect to the
  clock in the chamber. 
  The observed fraction of HH triggers is shown in Fig.~\ref{fig3}; it ranges
   from $\sim$65\% in a window of $\sim$10 ns to below 30\% in a window of less than  $\sim$5 ns;
    the curve has a periodicity of 25 ns.
 This structure is due to the finite frequency at which the signal wires are sampled in the BTI.

 In the LHC collisions, particles from the interaction region arrive in a muon
 station within an almost fixed time window, whose position depends on their average TOF.
If the arrival time of tracks, as seen by the DTLT electronics, were well defined with
  respect to the clock leading edge
 (as was the case for the Test Beam data mentioned above),
 one would simply set
 the delay such that the arrival time is on the plateau shown in Fig.~\ref{fig3} and a $\sim$5 ns accuracy would be sufficient. 
  However, this arrival time, called muon arrival time
   hereafter, has a spread
   given by the propagation of the signals from the track crossing
   point in the chamber to the front end electronics which can be as long as 10 ns in the $r$-$\phi$ plane; additional contributions
    are related to variations in TOF due to the bending in the magnetic field and
     to different path lengths due to the chamber dimensions ($\sim$4 ns). The resulting time window
      of 10--15~ns should overlap as much as possible
 with the plateau. Since this plateau has a width of $\sim$10 ns, we need a $\sim$1~ns precision
  to have an optimal efficiency of the HH triggers and the parent BX of all the tracks reaching a chamber uniquely    
        identified.
 
 \begin{figure}[h]
 %[htbp]
 \begin{center}
  \resizebox{10cm}{10cm}{\includegraphics[scale=0.75]{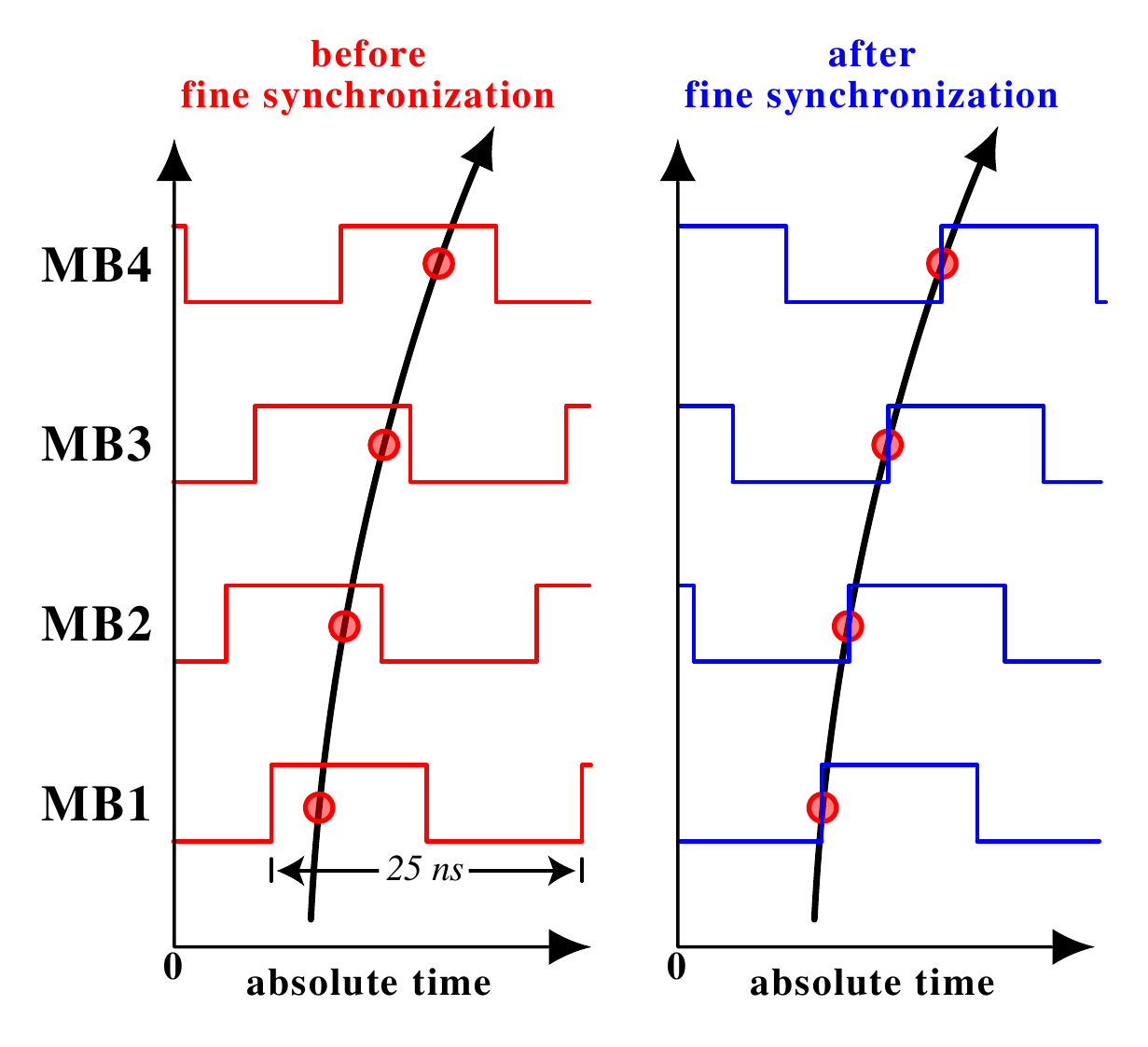}}
  \caption{ 
  Simplified visualization of a muon arrival time in the four stations of a sector of the DT barrel muon system
   as a function of the absolute time before and after "fine synchronization".
  The circle indicates the mean value of the arrival time of the  tracks crossing a chamber. The BX clock 
 leading edges of the different chambers
 are in a random position before fine synchronization (left plot). After fine synchronization (right plot) the phase
  of the clock edge in the chambers has 
 a specific value defined by maximizing the HH DTLT efficiency. 
 For simplicity, this value is assumed here to be zero for each chamber, when the phase is measured with respect to the leading edge of the clock.
   }
  \label{fig2}
 \end{center}
\end{figure}

The fine synchronization is illustrated for the four chambers of a sector in Fig.~\ref{fig2}.
 The absolute time is shown on the horizontal axis. The black curve represents the track of an outwards propagating muon 
coming from the
interaction region. The circles indicate the mean value of the muon 
arrival time in each station.
 Before synchronization (left plot), the BX clock edges of the different chambers
 are in a random position with respect to the absolute time due to differences
  in the distributed clock signals.  
  The fine synchronization consists in adjusting the clock in each chamber in order to
have the maximum efficiency of HH triggers. As shown in Fig.~\ref{fig3}, such maximum
efficiency is obtained when the muon arrival time in the chamber has a well defined phase with respect to the
local clock. The resulting setting is shown in the right part of the figure.
The spread of the arrival time of different tracks (due to different path lengths and signal time propagation) is not reported.
The time indicated by the circle represents the mean value of the arrival time of the tracks from the interaction point to the chamber
 of muons with a transverse momentum above a predefined value.
The fine synchronization is achieved adjusting the time phase between the sampling clock and the machine clock, by setting the fine delay
 (104 ps step) provided inside the TTCrx of each chamber.
  The change of the sampling time of
  the signals is equivalent to a modification of the reference time used in the
   track fit performed by the trigger electronics.  
 %
%%In Tab.~\ref{tab1} the expected TOFs of muons from the p-p interaction point to the DT stations are given.
%% The values include the spread due to the signal propagation along the wires which is about $\pm$5 ns 
%% for the wires measuring the $r$-$\phi$ position.  
%%
%% \begin{table}[htbp]
%% \begin{center}
%%  \resizebox{14cm}{!}{\includegraphics{OKtable_1}}
%%  \caption{TOF mean value and RMS for simulated muons coming from the  Interaction Point in DT chambers. }
%%  \label{tab1}
%% \end{center}
%%\end{table}
%tabella TOF

\section{Cosmic ray muon synchronization for CRUZET and CRAFT }

The CMS Collaboration conducted a month-long data-taking exercise,
known as the Cosmic Run At Four Tesla, during October-November 2008,
with the goal of commissioning the experiment for extended operation
\cite{CRAFTGeneral}.
With all the installed detector systems participating, CMS recorded 270 million
cosmic ray events with the solenoid at its nominal axial field
strength of 3.8~T.
Prior to CRAFT and during the final installation phase of the experiment, a
series of one week long commissioning exercises, to record cosmic ray events, took
place from May to September 2008. Progressively, an increasing fraction of the
experiment participated. These runs without magnetic field are known as the Cosmic RUn at ZEro Tesla
(CRUZET). Three hundred million
cosmic ray triggers were accumulated throughout such conditions.

 The DT muon barrel detector provided the trigger throughout the 
 cosmic ray data taking. The first level trigger signal was given by the coincidence, at the same BX, of at least two 
 DTLTs in different chambers in the same or neighboring sectors as described in Ref.~\cite{AlltriggCRuzet}. Since cosmic rays have no time relation with the CMS clock 
 and cross the detector with different angles and rates, a first coarse synchronization has been
  made in order to have, in almost all events, the same latency of the L1 signal independently of the geographical position 
  (chamber, sector, wheel) where the DTLTs were generated. 
  %This was performed taking into account the TOF of the cosmic muons, which generally cross the detector
  %from top to bottom, and increasing accordingly the trigger latency of the chambers of the top sectors.  
This was performed by aligning the BX of the DTLTs of the different chambers, more details can be found in Ref.~\cite{AlltriggCRuzet}.
%, taking into account the
% latency of the various trigger pipelines and the TOF of the cosmic muons, 
% which generally cross the detector from top to bottom.

%CFT-09-013 line 411 ...The local trigger synchronisation was specifically adjusted for cosmic ray triggers. Cosmic ray
% muons generally cross the detector from top to bottom, so the system was synchronised to take this into account.
%  The trigger latency of the chambers of the top sectors was increased accounting for a maximum time
% of flight to the bottom chambers of about 50 ns, or 2 BX. Thus, when a single muon crosses two
% sides of the detector, two segments from different detector regions are sent to the DTTF in the
% majority of cases for the same BX so that the DTTF system sends two muon track candidates to
% the GMT at the same BX 
%CFT-09-025 line 227....  Therefore, at the LHC start-up, after the initial coarse synchroniza
% tion, obtained from the adjustment of TOF from pp interaction to the coarse synchronization
%used for the CRAFT runs, the DT chambers are expected to trigger on muons from p-p inter
% actions with a reasonable efficiency even before any fine synchronization.

For the fine synchronization performed during CRUZET and CRAFT, specific data were collected triggering with only one chamber
 in each of the top sectors (sectors 3, 4, 5 of each wheel, see e.g. Ref.~\cite{DTtriggCRAFT}).
 The BX number was assigned to the event by the DTLT of the chamber giving the L1 trigger,
 whereas the BX distribution 
 of HH triggers in chambers not contributing to the L1 was used to provide the 
direct measurement of the relative clock phase. 

 \begin{figure}[htbp]
 \begin{center}
  \resizebox{10cm}{20cm}{\includegraphics{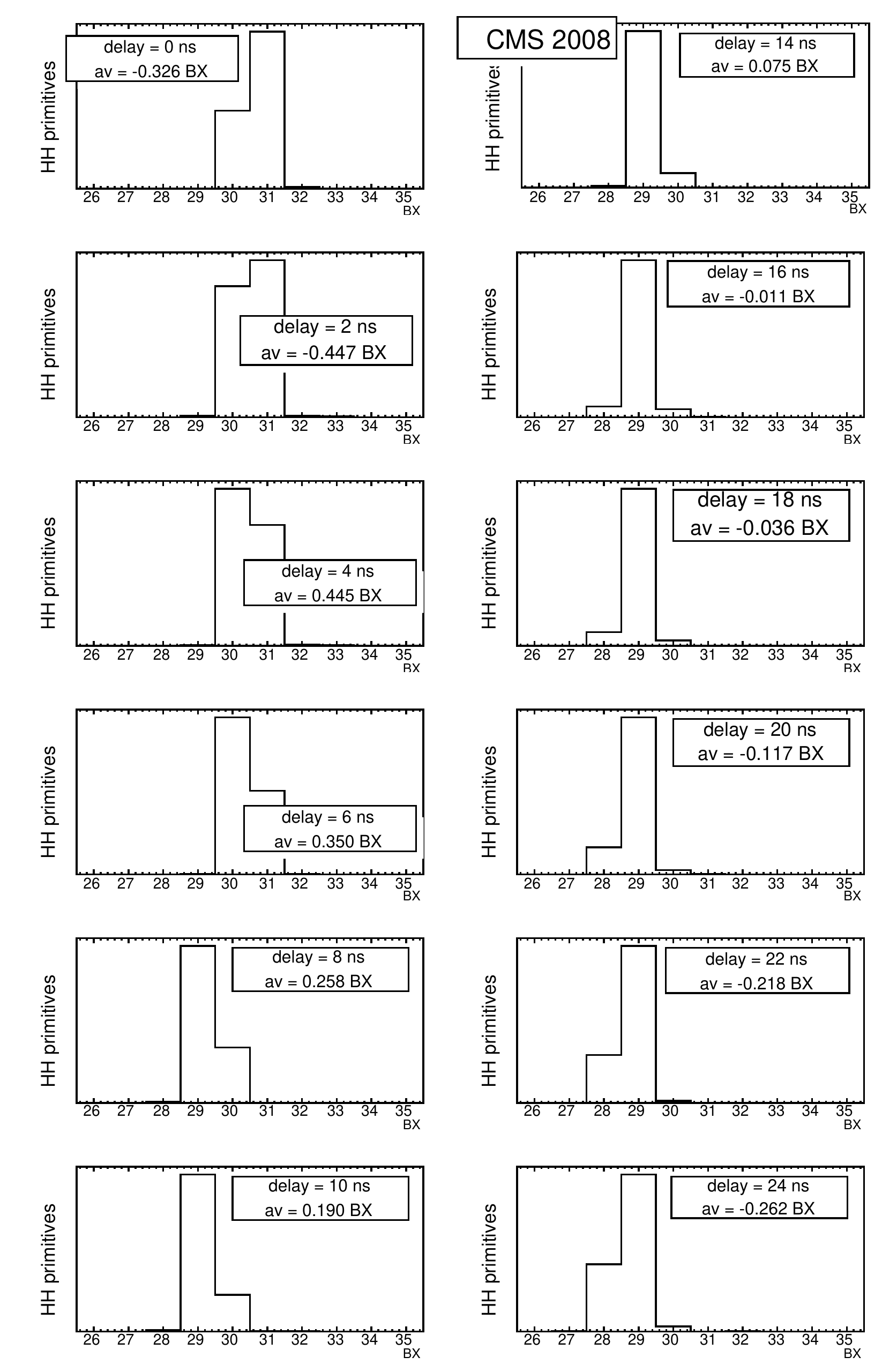}}
  \caption{BX distribution of high quality (HH) trigger primitives in one chamber. Data are from the MB4 chamber of
    sector 11 of wheel $-$1, with a  L1 signal taken from station MB3 only; there are about 1500 events for each
     delay setting. The histograms are scaled to a common maximum height. The observed average BX number with respect to
      the BX of the peak is indicated for each delay. For example, when no delay is applied, the bin BX = 31 has the largest statistics
and the mean value of the BX distribution is 30.674, resulting in $av = -0.326$ BX.
The shift of one BX for the
   plots with delay above 6 ns is due to a change of the trigger latency. }
  \label{fig4}
 \end{center}
\end{figure}

 The periodic dependence of the BX distribution shape with respect to the TTCrx fine delay was
  inspected using a dedicated set of 12 runs.
 In each run, the DTLT data of a whole sector were read out using the L1 strobe generated by the MB3 station
  only. Run-by-run the clock phases of the other three stations were moved in steps of 2 ns thus allowing 
   a complete phase scan. The MB3 TTCrx delay was left fixed at 12.5 ns. The BX distribution of DTLT trigger 
primitives with HH quality is shown
    in Fig.~\ref{fig4} for one chamber of the MB4 station and for different values of its TTCrx delay.
The BX distribution changes when shifting the phase, passing from a sharp peak, through intermediate
 stages, where events are shared in two contiguous BXs, to a sharp peak again, shifted by one BX with respect to
  the previous one. As an indicator of "unique BX identification", 
  hence of a good fine synchronization,
   we used the difference, $av$, between the average of the BX
distribution and the BX with the maximum number of 
entries\footnote{The BX of the MB3 chamber that provided the trigger signal could have been chosen as reference.
The choice of using instead the BX peak value of each chamber as their own reference
 makes more apparent the specific scope of the local trigger fine synchronization with respect to
  the initial coarse synchronization.
The  initial coarse synchronization
  equalizes  the cable paths chamber-to-chamber in order to give constant latency for the L1. While the coarse synchronization aligns in time
     the BX peak values,  the fine synchronization aims at optimizing the BX distribution.}.
     This is a convenient variable since it is independent of the latency actually used
      and is proper to each chamber.   Any muon trigger,
      after coarse synchronization, could have been used for collecting the data: operationally we have
       chosen to trigger on a single chamber for
     simplicity, and MB3 chambers were the obvious choice since they are the largest ones, 
     giving the largest rate of cosmic ray muons.
     
 \begin{figure}[htbp]
 \begin{center}
  \resizebox{12.5cm}{!}{\includegraphics[angle=90]{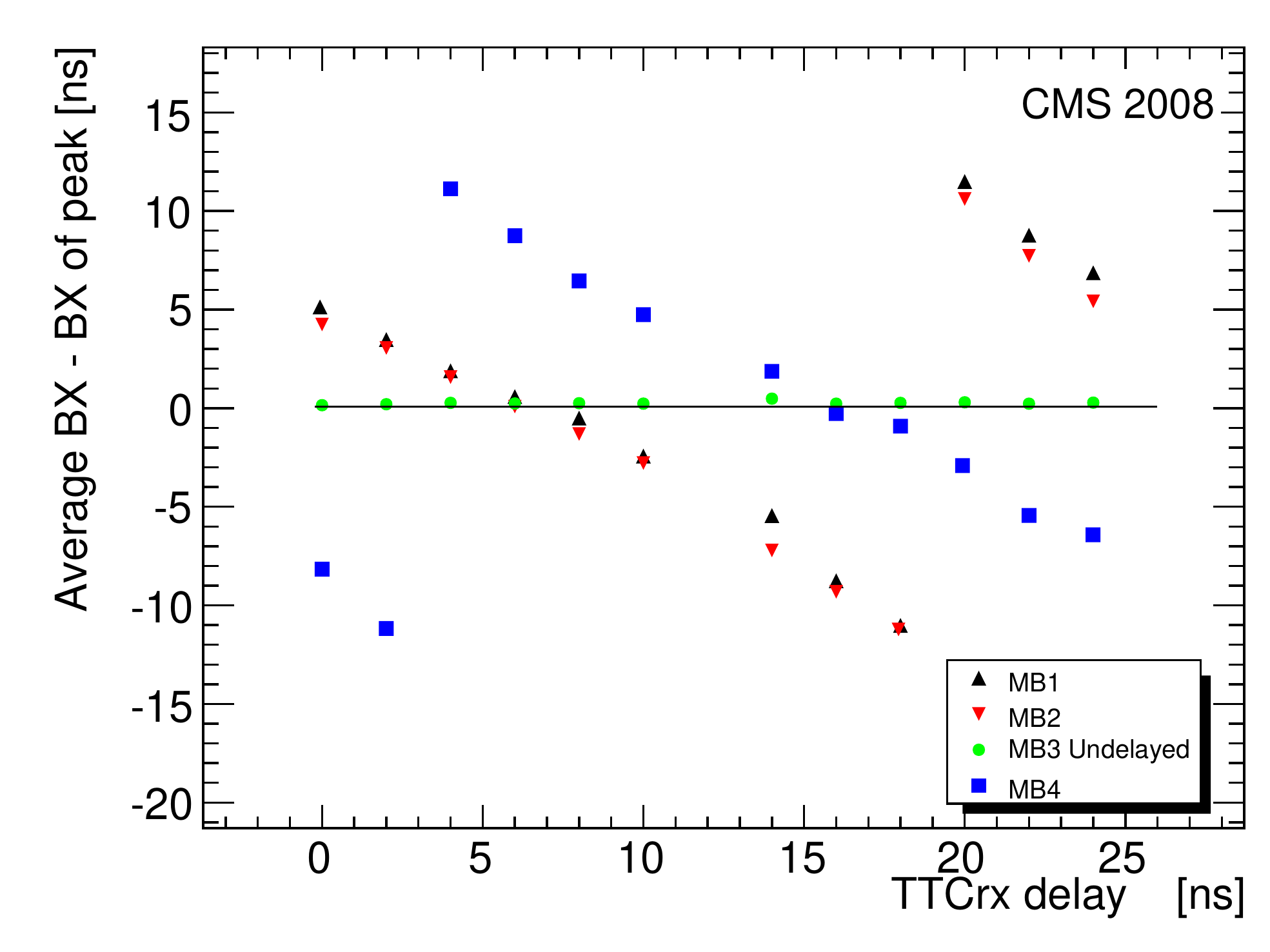}}
  \caption{Average BX time for stations MB1, MB2 and MB4, measured with respect to the time of the BX bin
   with the largest number of entries,
   as a function of the applied TTCrx delay. Data are from sector 11 of wheel 1, with a L1 signal from station MB3 only,
    which is taken as reference and had a constant TTCrx delay set at 12.5 ns. The average BX time for MB3 is consistent with zero as 
    expected. There are about 1500 events for each delay setting.}
  \label{fig5}
 \end{center}
\end{figure}
\begin{figure}[htbp]
 \begin{center}
  \resizebox{12cm}{!}{\includegraphics[angle=90]{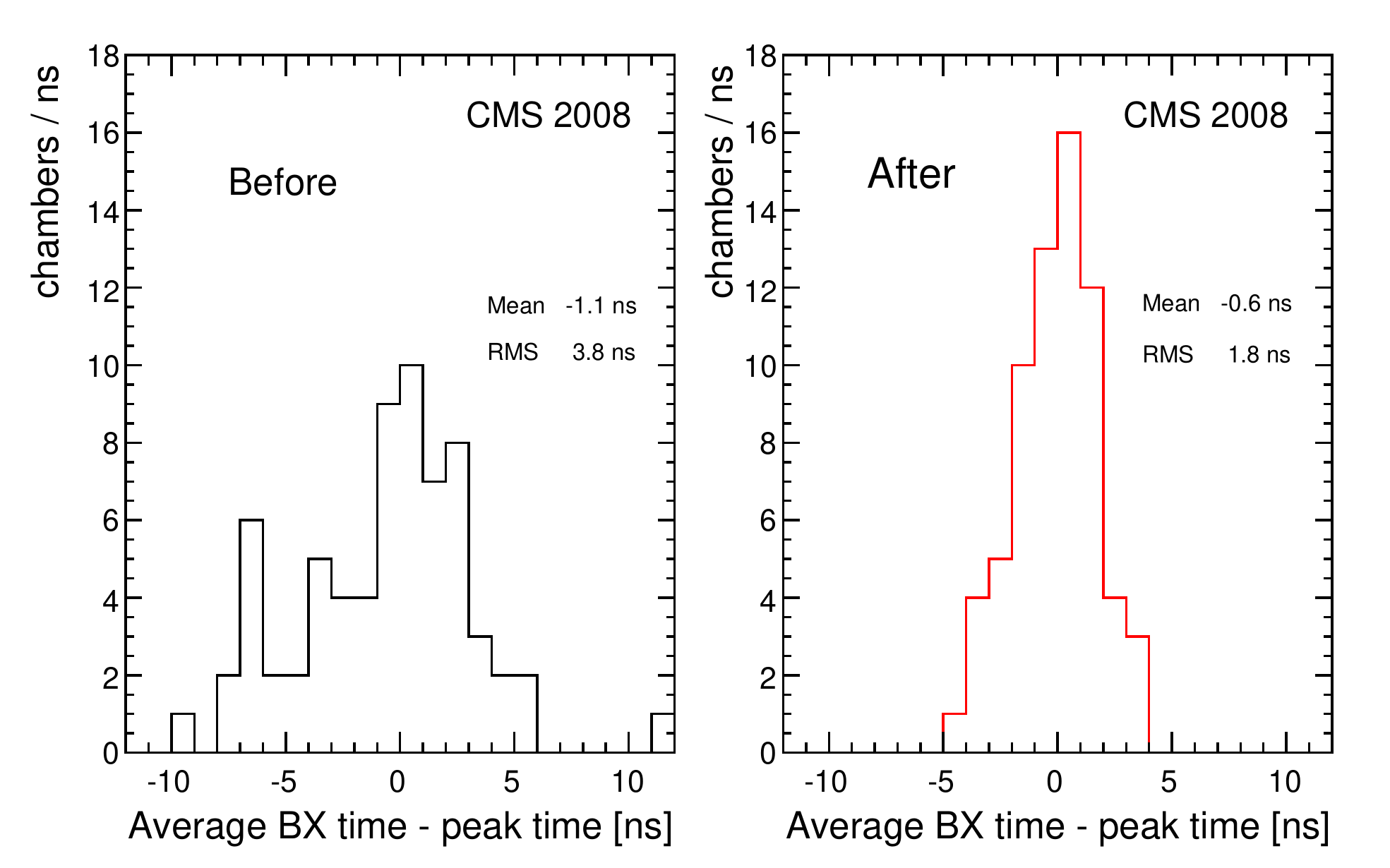}}
  \caption{Distribution of the average BX time with respect to the time of the BX bin with the largest number of entries 
  before (left plot) 
  and after (right plot) the delay adjustment for all chambers of sectors 3, 4, 5, 9, 10 and 11 of the wheels 0, 1 and 2.
   In these data, sector 9 of wheel +2 was not read out.}
  \label{fig6}
 \end{center}
\end{figure}
 The distribution of $av$ in ns is plotted in Fig.~\ref{fig5} for MB1, MB2 and MB4 as a function of their
  local TTCrx delay. As expected, it ranges
     from $-$0.5 BX to 0.5 BX, i.e.\ from $-$12.5~ns to +12.5~ns, being zero by definition at the optimal phase
value. These plots show that the average BX time depends linearly on the TTCrx fine delay, 
the deviations from linearity being less than 2 ns. Having verified this linear dependence 
      through this preliminary analysis, the fine 
      synchronization  for CRAFT was performed measuring the time of the average BX with the data collected
       in each chamber 
      and shifting their TTCrx fine delay accordingly.    
As an example, Fig.~\ref{fig6} shows the distribution of  $av$ values for a subset of chambers before and
 after the TTCrx delay adjustment,
triggering with all studied sectors. The distribution after the delay adjustment is narrower.
 
Throughout the CRAFT run, the DT Track Finder
 provided a stable muon trigger rate of $\sim$250~Hz as described in Ref.~\cite{AlltriggCRuzet}. 
 The stability of the synchronization was monitored and verified continuously during the data taking. It has been
  checked using the mean value of the BX
   distribution of primitives with HH quality in each chamber.

The method described here, used for the fine synchronization of cosmic ray data taking, can only be applied when the
 arrival time of the tracks spans over a time range of at least one BX period. Therefore, the method cannot be applied
  successfully during LHC operation, where the arrival times of muons coming from the interaction region are confined
   within a range of 10--15 ns.
   %%, as   reported in Tab. ~\ref{tab1}. 

\section{DT Local Trigger efficiency as a function of time and of clock phase with cosmic ray data}

As explained in Section 2, the DT local trigger performance is optimal when the muon arrival
 time in a chamber, as seen by the DTLT electronics, has a well defined phase with respect to the local
clock in the chamber itself. Cosmic rays, having a flat time distribution,
 allow this optimal phase to be determined, by measuring the 
 HH trigger efficiency for each of the 250 DT chambers as a function of the muon arrival time. 
The result, i.e.\ the optimal phase, is an intrinsic property of each 
chamber and is independent of the actual value of the TTCrx delay.

The arrival time of a track segment
%segment
 is reconstructed 
from the readout data of the DT chambers as explained in Ref.~\cite{LNLt0segment}. Close hits are grouped together by a pattern
 recognition algorithm. The track direction and
position are determined assuming a straight line trajectory.
In the fit, a common displacement of the hits from the wire position is treated as a free parameter.
 Assuming a constant drift velocity,
 this common displacement corresponds to a shift in the time of the track.
This reconstructed time 
shift 
has a resolution of about 3 ns \cite{DTLocalRecoCRAFT}.

\begin{figure}[htbp]
 \begin{center}
  \resizebox{16cm}{!}{\includegraphics[angle=90]{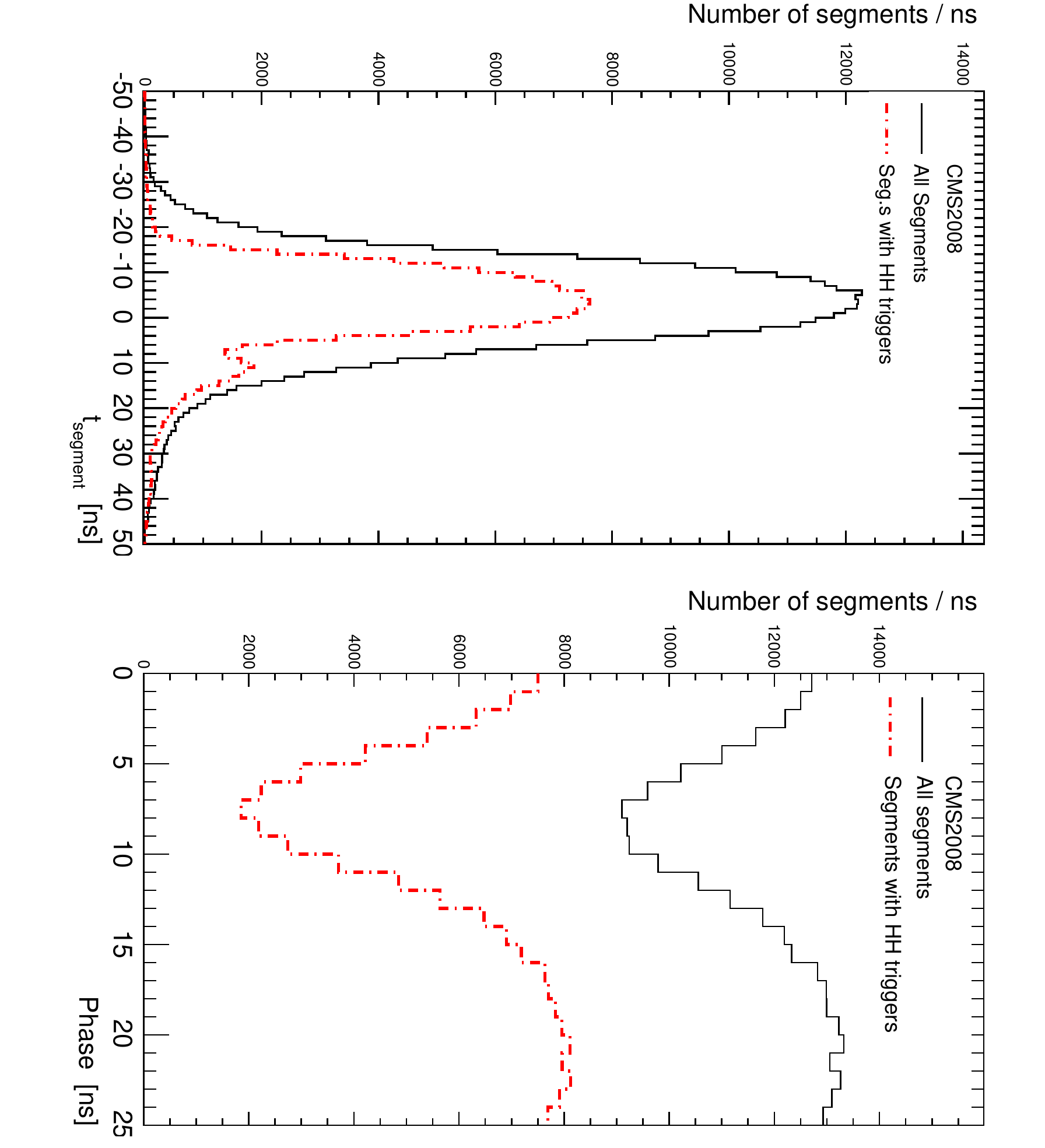}}
  \caption{ Left: time distribution of track
segments in one chamber with (dash-dotted histogram) and without (continuous histogram)
the HH quality requirement.    The minimum in the distribution of HH triggers locates 
the boundary between two BXs.
 Right: the same distributions folded into a 25 ns interval.
 % $Phase$ is the phase of $t_\textrm{segment}$ with respect to the 25 ns clock.
 }
  \label{fig7}
 \end{center}
\end{figure}

For the purpose of this study, the reconstructed time shift ($t_\textrm{segment}$)
is measured with respect to the leading edge of the clock of the chamber. 
The resulting distribution for all reconstructed segments
in the $r$-$\phi$  projection
 is shown  in the left plot of Fig.~\ref{fig7} for one chamber. The
 subset of the tracks in this chamber with an associated HH trigger is shown in the same plot.
   The second local maximum, visible in the distribution of this subset, is associated with the fraction of events
   triggered in the chamber at a BX contiguous to the one defined by the L1 trigger.
Given the periodical behavior of the efficiency, these distributions were folded in the central
 25 ns interval, as shown in the right plot, in order to increase statistics.
 Hereafter, the $t_\textrm{segment}$ folded in 25 ns interval is referred to as the "$Phase$".
  
 \begin{figure}[htbp]
 \begin{center}
 \resizebox{16cm}{!}{\includegraphics[angle=90]{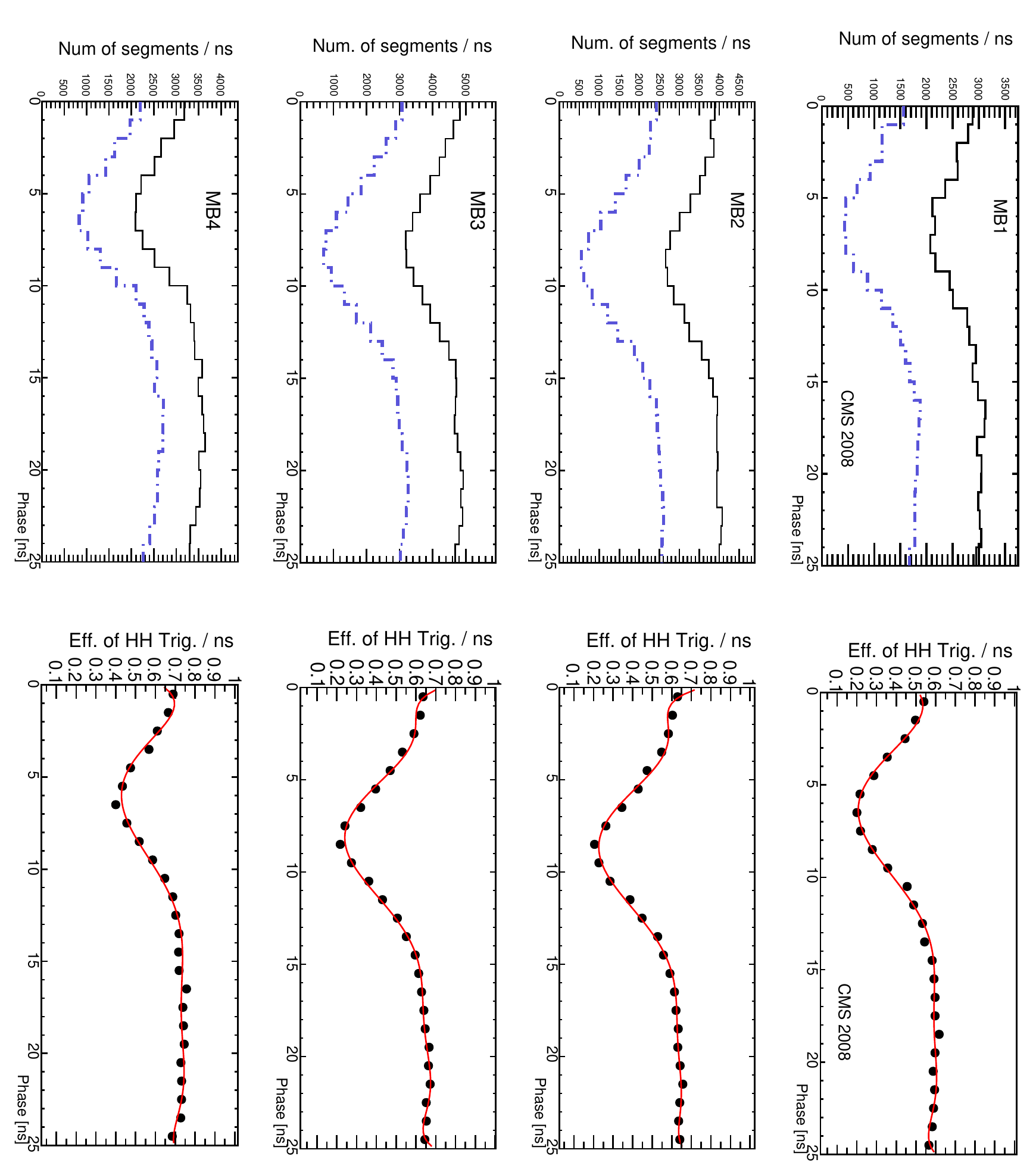}}
  \caption{Left: $Phase$ distribution for all reconstructed segments (continuous histogram) and
   for segments with an associated HH trigger (dash-dotted histogram)
       of the four chambers of one sector. Right:
   corresponding HH efficiencies as a function of $Phase$, with a polynomial fit for each chamber.}
  \label{fig8}
 \end{center}
\end{figure}

\begin{figure}[htbp]
 \begin{center}
  \resizebox{16cm}{!}{\includegraphics[angle=90]{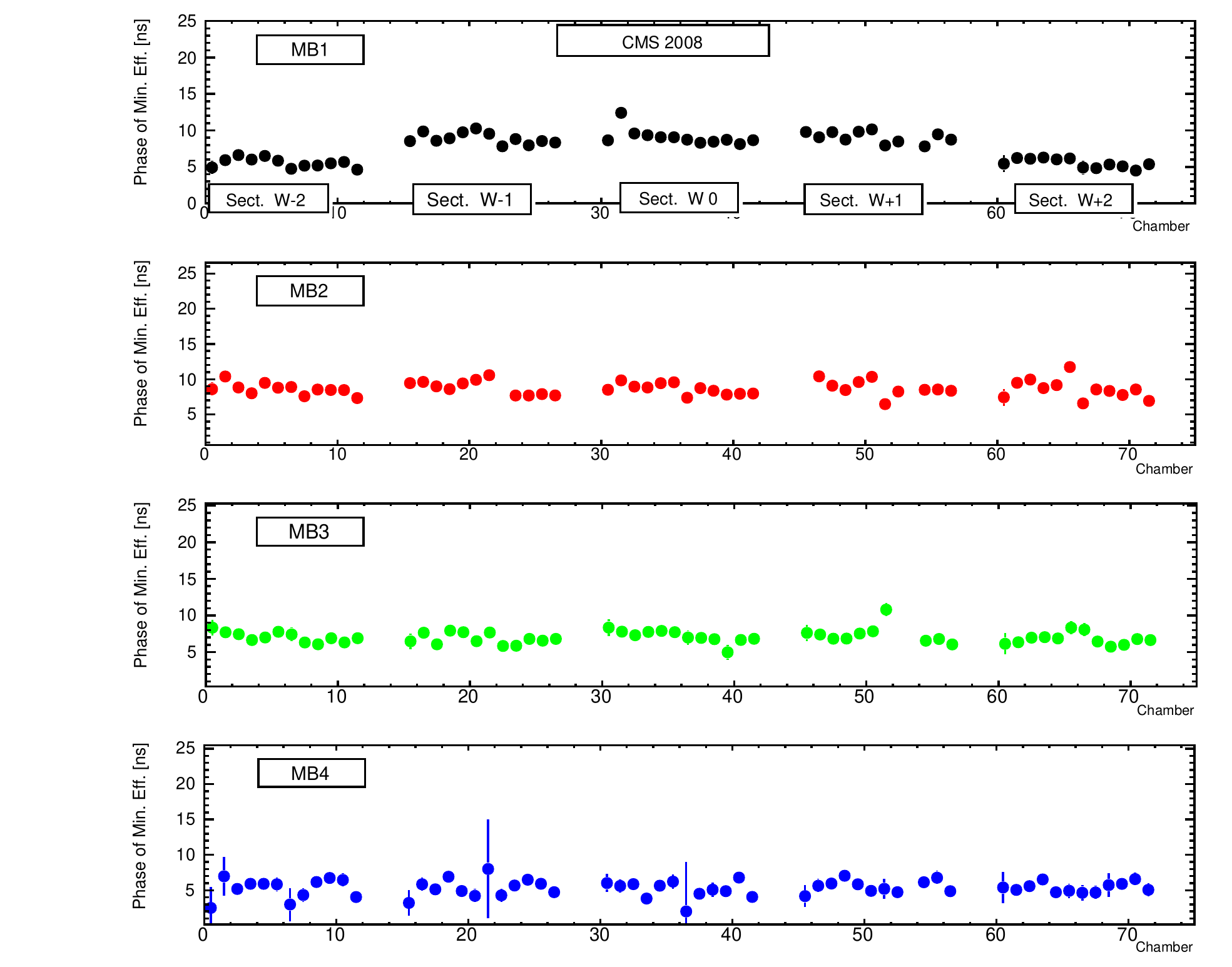}}
  \caption{$Phase$ of the minimum of HH quality DTLT efficiency.
   On the $x$ axis, each entry corresponds to one chamber ordered by sector (1 to 12), 
 and by wheel (-2 to +2). Some chambers are missing for which the triggers primitives were not read out, 
 or a high voltage channel was disconnected.
 %     No result is shown for the chambers of sector 9 of wheel +1  
 %  for which the trigger primitives were not readout (even if generated) and for the MB3 chamber
 %  of sector 8 of wheel +1 which had the HV of one layer disconnected. 
   }
  \label{fig9}
 \end{center}
\end{figure}

 \begin{figure}[htbp]
 \begin{center}
  \resizebox{15cm}{10cm}{\includegraphics[angle=90]{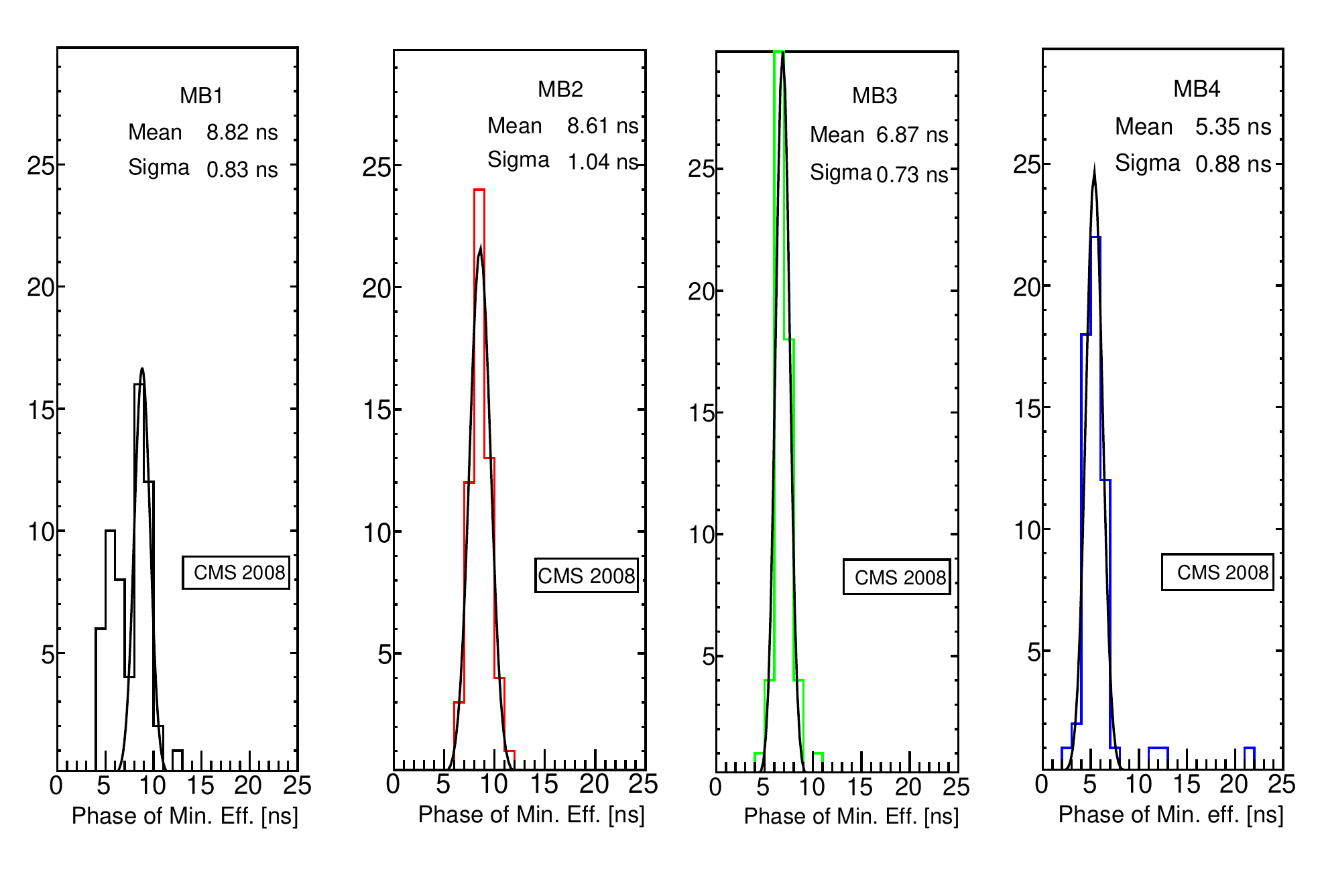}}
  \caption{Distributions of $Phase_\textrm{min}$ 
   for each chamber type, from left to right for MB1, MB2, MB3 and MB4.
     In the MB1 chambers of wheels +2 and $-$2, the $Phase_\textrm{min}$
  is different from that in the other MB1 chambers because the drift velocity is not the same, due to the stray magnetic field.
}
  \label{fig10}
 \end{center}
\end{figure}

In Fig.~\ref{fig8}, for each of the four chambers of one sector, the $Phase$ of all tracks and of the ones with 
an associated HH trigger are shown on the left
 side; the HH trigger efficiency, i.e\ the ratio of the number of tracks with HH quality to the total number of tracks in the chamber,
 is shown on the right side. The plots show that the DTLT
  efficiency as
  a function of $Phase$ is maximal over a plateau $\sim$10 ns wide
   and has a minimum in a region $\sim$5 ns wide (similar to the one measured with the bunched muon test beam and
    reported in Fig.~\ref{fig3}).
     These curves were fitted with a polynomial
     in order to find the $Phase$ of the minimum of efficiency ($Phase_\textrm{min}$) in each chamber,
      since the position of the minimum is much better defined than that of the maximum. The optimal phase is hence 
defined as $Phase_\textrm{min}$ shifted by 12.5 ns.    
 \begin{figure}[htbp]
 \begin{center}
  \resizebox{16cm}{10cm}{\includegraphics[angle=90]{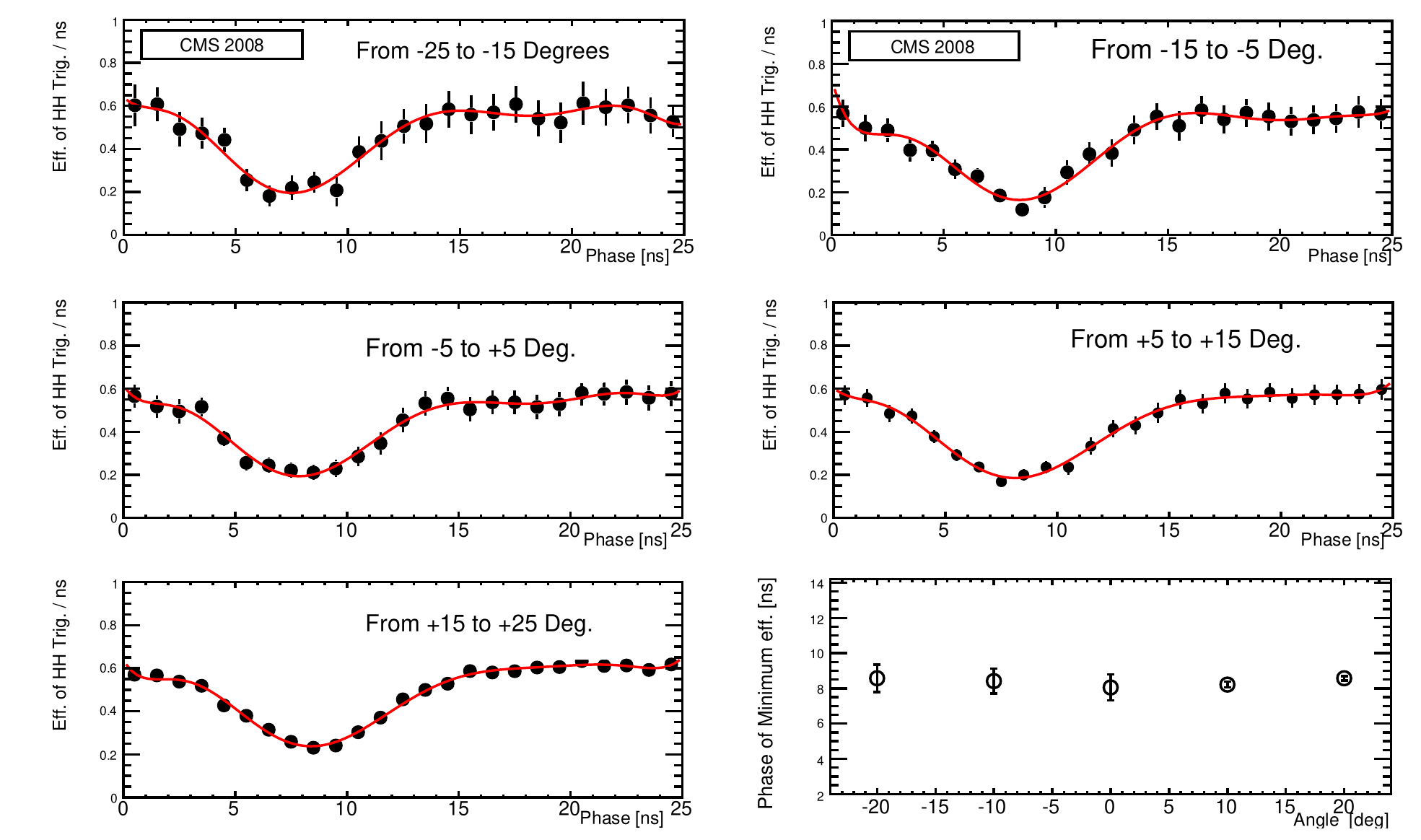}}
  \caption{DTLT of HH quality efficiency as a function of $Phase$ for five ranges of the track incidence angle:
   $[-25,-15], [-15, -5], [-5, +5], [+5, +15], [+15, +25]$ degrees. 
  The $Phase$ of the minimum efficiency has no dependence on the angle, as shown in the bottom right plot. }
  \label{fig11}
 \end{center}
\end{figure}
Figure ~\ref{fig9} shows the resulting $Phase_\textrm{min}$ for all DT chambers
 with the local trigger data from one CRAFT run.
Results from other runs agree, within statistical and systematic errors.
 %On the $x$ axis, each entry corresponds to one chamber ordered by sector (1 to 12), 
 %and by wheel ($-$2 to +2).
 The errors on $Phase_\textrm{min}$
of the chambers in the vertical sectors (namely sectors 1 and 7) are much larger than the others
   because of the poorer reconstruction efficiency and accuracy caused by the large angle of cosmic ray 
   tracks in these sectors. 
   Systematic errors, caused by the calibration parameters \cite{DTcalibCRAFT} used for referring the time to the
    clock leading edge, are not reported. In the
   chambers of sectors oriented close to the horizontal direction
   % horizontally positioned 
   (sectors 3, 4, 5, and 9, 10, 11) these errors are lower than  1 ns;
    in the
   chambers of other sectors they can be as large as a few ns.
The results are summarized in Fig.~\ref{fig10}, where the distributions of $Phase_\textrm{min}$,
  shown individually in Fig.~\ref{fig9},
   are plotted for the four station types separately.
The $Phase_\textrm{min}$ 
is the same within a few ns for
 each chamber type (a Gaussian fit to the distribution of $Phase_\textrm{min}$ returns a $\sigma$
smaller than 1 ns).  This is a consequence of 
similarities in the construction of the chambers and their operating conditions, and can be
used to  reduce
the systematic bias which affects the measurements in the vertical sectors. 
  It should be noted that in the MB1 chambers of wheels +2 and $-$2, $Phase_\textrm{min}$
  is different from that in the other MB1 chambers. This is an effect of the drift velocity in these chambers,
   which is different from the
  others due to the stray magnetic field,
  as already observed in  previous measurements \cite{ MTCCvdrift,DTtest1}. 
 No dependence of $Phase_\textrm{min}$ on the track incidence angle was observed in the range of $-$25 to +25 degrees,
 corresponding to the angular interval expected for
  high transverse momentum muons produced at LHC. Figure~\ref{fig11} shows the efficiency curves as a function of
 $Phase$, measured for five ranges of angles. The plot on the bottom right side shows
 $Phase_\textrm{min}$ as a function of the angle. 

% \begin{figure}[htbp]
%  \begin{center}
%   \resizebox{14cm}{!}{\includegraphics[angle=90]{OKFigure_12}}
%   \caption{Distribution of $t_\textrm{segment}$ as computed for simulated muon from $pp \to \mu X$  at the LHC.
% In the simulation the muon barrel chambers are already calibrated 
%   (mean value set to zero). }
%    \label{fig12}
%  \end{center}
% \end{figure}

\section{Fine synchronization at the LHC start-up}

The study reported in Ref.~\cite{DTtriggCRAFT} shows that, summing over all trigger qualities (HH, HL, LL),
 the efficiency of DTLT primitives is
 above $\sim$60\% for any value of the local clock phase, with an 
average value
   of $\sim$80\% .
Therefore, at the LHC start-up, after the initial coarse synchronization\footnote{
To achieve this coarse synchronization,  the TTCrx delays of the different chambers
can be shifted from the values used during CRAFT, by an amount which can be determined
from the TOF difference between cosmic muons and muons coming from p-p interactions.}, the DT chambers 
	  are expected to trigger on muons from p-p interactions with a reasonable efficiency even before
	   any fine synchronization.
	    Triggers from other detectors will be used as well.
	    % Provided that 
	    Since the latency accuracy of such triggers will be in
	     a range of two BXs, 
	     % given the knowleadge of TOF relations and the synchronization experience from cosmic runs,
	      the segment reconstruction and the related time measurement in the DT chambers will be carried out correctly.
%\begin{figure}[htbp]
\begin{figure}[htbp]
 \begin{center}
  \resizebox{12cm}{!}{\includegraphics{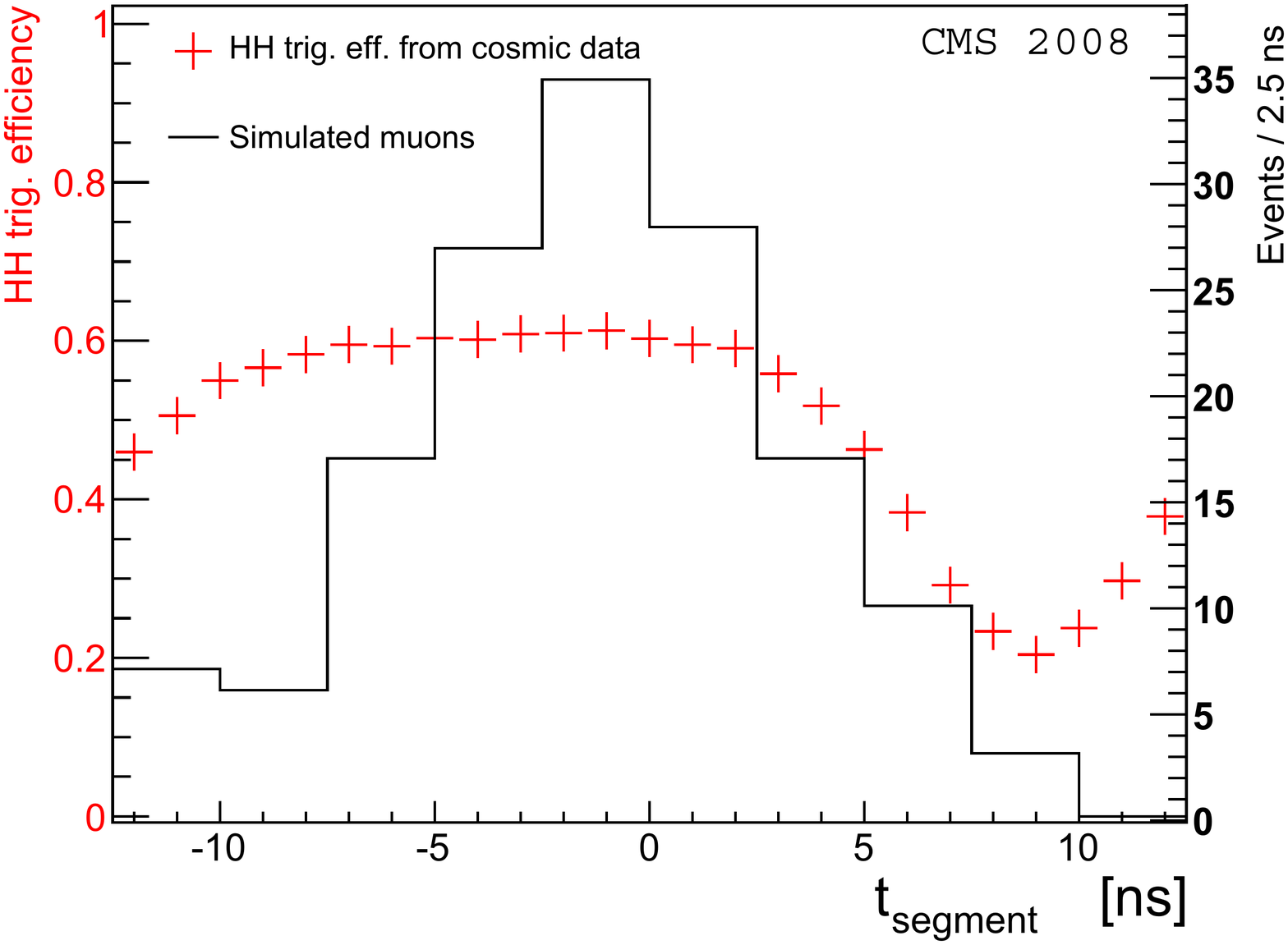}}
  \caption{Distribution of $t_\textrm{segment}$ for simulated muons from pp~$\to\mu$X reaching a DT chamber
 (histogram, units on the right side axis), together with the corresponding efficiency curve,
  as measured with cosmic rays (crosses).
   %The time range shown is limited to the range of the reconstructed time of the simulated muons.
    }
  \label{fig13}
 \end{center}
\end{figure}
At the LHC start-up, it will therefore be possible to produce the $Phase$ distribution
 of tracks 
 in any chamber. This distribution is expected to be much narrower than the one obtained with cosmic rays,
  and to peak at a fixed position. In order to get the maximum HH trigger efficiency, the peak has to be moved 
  to the
   position of maximum efficiency already obtained with cosmic rays as described in Section 4.
The fine synchronization will consist in the application, chamber by chamber,
 of the optimal delay, obtained as the difference between the phase peak at start-up and the phase of the maximum
  of the efficiency. 
  
Muons from p-p interactions have an arrival time in a chamber with a spread of ~10--15~ns,
Because the tracks are not distributed randomly with respect to the clock, the mean value 
 can be determined with an accuracy
 of $\sim$1 ns with only a few hundred tracks per sector coming from the
 interaction point. 
The time needed to collect a few hundred muons per chamber should be very short since the acquisition
 will take place in
 all chambers concurrently, without any need to upload specific parameters to the DTLT electronics, as would be the case
  for a scan of the TTCrx delays. Moreover, the procedure can be applied to data collected either centrally
   or with the DT local
   data acquisition, with any source of the L1 trigger. 
The rate of background cosmic ray events is expected to be negligible, reduced by a factor greater than $10^3$ with respect to the $\sim$250 Hz 
measured in the CRAFT runs, by requiring pointing tracks with transverse momentum larger than $\sim$10 GeV/c, at trigger and
at reconstruction level.  

In Fig.~\ref{fig13}, the distribution of $t_\textrm{segment}$ for muons coming from a sample of simulated pp~$\to\mu$X events,
 from LHC collision,
  is shown for one chamber. Even though at the LHC start-up we may expect distributions with different mean values,
    their widths should be similar to the one reported. The  $t_\textrm{segment}$ distribution for these muons will
     be narrow compared to the width
  of the maximum efficiency plateau; this confirms that the precision of the measurements presented here is enough to ensure a good synchronization of the DT
  trigger for muons produced in p-p collisions.

\section{Summary}

The random arrival time of cosmic ray muons allowed several synchronization aspects
 of the Drift Tube Local Trigger to be studied.
 The method used for the synchronization for cosmic ray data has been presented. The stability of the synchronization
  over a period of several months has been verified.
  
  A new method to achieve the DTLT synchronization at the LHC was presented. Cosmic ray events, which span the whole 25 ns clock cycle,
   were used to
  measure the efficiency curve of high quality DT local trigger primitives in each chamber, as a function of the phase of the
  reconstructed muon arrival time with respect to the local clock edge. The optimal phase between the muon signal and the local clock,
   ensuring maximal
  efficiency and unique BX identification, was determined for each chamber.
  For most chambers, a precision better than 1 ns was reached. For chambers belonging to the same wheel and station, the optimal
  phases are similar, 
  because the chambers are of identical design and are operated under the same conditions.
  %resulting from similarities in their construction and operating conditions.
   This can be used to reduce
   the systematic bias which affects
  the measurements for the chambers where the direction of cosmic ray muons is very different from that of muons from p-p
collisions.

Since these optimal phases have now been determined, a few hundred tracks per chamber will be sufficient to
 perform the synchronization of 
the DT trigger very quickly at the LHC start-up. Compared to the method envisaged originally, all the
 required data can be collected with a single running configuration.
%The method is anyway independent of the method foreseen in previous studies based on many runs scanning the clock range;
%at LHC both can be performed.

\section*{Acknowledgments}

We thank the technical and administrative staff at CERN and other CMS
 Institutes, and acknowledge support from: FMSR (Austria);
  FNRS and FWO (Belgium); CNPq, CAPES, FAPERJ, and FAPESP (Brazil);
   MES (Bulgaria); CERN; CAS, MoST, and NSFC (China); COLCIENCIAS (Colombia);
    MSES (Croatia); RPF (Cyprus); Academy of Sciences and NICPB (Estonia);
     Academy of Finland, ME, and HIP (Finland); CEA and CNRS/IN2P3 (France);
      BMBF, DFG, and HGF (Germany); GSRT (Greece); OTKA and NKTH (Hungary);
       DAE and DST (India); IPM (Iran); SFI (Ireland); INFN (Italy);
        NRF (Korea); LAS (Lithuania);
	 CINVESTAV, CONACYT, SEP, and UASLP-FAI (Mexico);
	  PAEC (Pakistan); SCSR (Poland); FCT (Portugal);
	   JINR (Armenia, Belarus, Georgia, Ukraine, Uzbekistan);
	    MST and MAE (Russia); MSTDS (Serbia); MICINN and CPAN (Spain);
	     Swiss Funding Agencies (Switzerland); NSC (Taipei);
	      TUBITAK and TAEK (Turkey); STFC (United Kingdom);
	       DOE and NSF (USA). 
	       Individuals have received support from the Marie-Curie IEF program (European Union);
	        the Leventis Foundation; the A. P. Sloan Foundation;
		 and the Alexander von Humboldt Foundation. 

\bibliography{auto_generated}   % will be created by the tdr script.
\clearpage
\appendix

\cleardoublepage\appendix\section{The CMS Collaboration \label{app:collab}}\begin{sloppypar}\hyphenpenalty=500\textbf{Yerevan Physics Institute,  Yerevan,  Armenia}\\*[0pt]
S.~Chatrchyan, V.~Khachatryan, A.M.~Sirunyan
\vskip\cmsinstskip
\textbf{Institut f\"{u}r Hochenergiephysik der OeAW,  Wien,  Austria}\\*[0pt]
W.~Adam, B.~Arnold, H.~Bergauer, T.~Bergauer, M.~Dragicevic, M.~Eichberger, J.~Er\"{o}
% ...
\vskip\cmsinstskip
% ...

\end{sloppypar}
\end{document}